\pdfoutput=1

\documentclass{article}

\usepackage{graphicx}
\usepackage{amsmath,amssymb,amsfonts}
\usepackage{color}
\usepackage{xr}
\usepackage{float}
\usepackage{bm}
\usepackage[colorlinks,linkcolor=blue,citecolor=blue]{hyperref}
\usepackage{color}

\begin{document}
{\flushleft\Large{\bf Evaluating genetic drift in time-series evolutionary analysis}}\\[2ex]

Nuno R. Nen\'{e}$^{1}$,
Ville Mustonen$^{2}$ and
Christopher J. R. Illingworth$^{1}$

{\flushleft
${}^1$Department of Genetics, University of Cambridge, Cambridge, UK
${}^2$Wellcome Trust Sanger Institute, Wellcome Trust Genome Campus, Hinxton, Cambridge CB10 1SA, UK
${}^3$ E-mail: cjri2@cam.ac.uk
}

\begin{abstract}

The Wright-Fisher model is the most popular population model for describing the behaviour of evolutionary systems with a finite population size.  Approximations to the model have commonly been used for the analysis of time-resolved genome sequence data, but the model itself has rarely been tested against genomic data.  Here, we evaluate the extent to which it can be inferred as the correct model given experimental data.  Given genome-wide data from an evolutionary experiment we validate the Wright-Fisher model as the better model for variance in a finite population in contrast to a Gaussian model of allele frequency propagation.  However, we note a range of circumstances under which the Wright-Fisher model cannot be correctly identified.  We discuss the potential for more rapid approximations to the Wright-Fisher model.

\end{abstract}




\section{Introduction}\label{Intro}


Rapid advances in high-throughput methodologies have enabled the collection of rich time-series from experimental evolution studies. These typically address the effects of environmental conditions on adaptation stemming from de novo mutations \cite{Barrick2013}, initial variance induced by a genetic cross \cite{Culleton2005,Mancera2008,Bergstrom2014} or simply from the standing variation characterizing a polymorphic starting population \cite{Schloetterer2014}. Sequencing the emerging populations during these types of experiments allows for identification of molecular aspects behind the species' reproductive success.\\


Despite advances in the field, a challenge remains regarding the optimal approach for identifying loci under selection given time-resolved genomic data.  Due to linkage disequilibrium, selection at a single locus can lead to changes in allele frequencies across multiple loci~\cite{Hill:1966fe}, confounding single-locus approaches to the inference of selection ~\cite{Illingworth2011}.  Further, in smaller populations, genetic drift may have a significant impact upon allele frequencies, such that the influence of selection must be distinguished from stochastic effects, arising from both propagation and sampling \cite{Jorde2007,Charlesworth2009,Jonas2016}. \\

A variety of methods have been proposed for inferring selection under genetic drift, utilising the Wright-Fisher drift model for forward propagation \cite{Ewens2012}, approximations to the Wright-Fisher model \cite{Waxman2011,Feder2014,Lacerda2014,Terhorst2015,Topa2015}, its diffusion limit~\cite{Bollback2008} and respective spectral decomposition approaches \cite{Song2012,Steinruecken2014}, or effective simulation methods \cite{Foll2015,Malaspinas2016}.  However, while the Wright-Fisher model has become the standard approach to representing genetic drift, it is built upon certain modelling assumptions, including the replacement of the entire population in successive generations.  As such, other models may in some respects provide a better fit to the dynamics observed in evolutionary experiments \cite{Der2011}.  Experimental demonstrations intended to validate the Wright-Fisher model have suffered from limitations in the extent of data available for analysis~\cite{Der2011,Buri1956}. \\

Here we evaluate the extent to which a Wright-Fisher model of genetic drift can be inferred from data collected from evolutionary experiments, contrasting it with a model of Gaussian diffusion.  The Gaussian model at first sight differs greatly from the Wright-Fisher model, lacking frequency dependent variance, albeit we note that, when compounded with the effect of finite sampling, frequency-dependent variance does arise in the Gaussian model.  A further contrast is noted in the computational efficiency of the algorithms; the Gaussian model is analytically solvable, allowing for rapid evaluation, whereas the Wright-Fisher model is more computationally intensive.  Considering simulated allele frequency data we note that inference of a Wright-Fisher model is not always possible, with various parameters contributing with different degrees towards greater model identifiability.  Considering a large dataset from evolutionary experiments conducted in {\it Drosophila melanogaster} ~\cite{OROZCO-terWENGEL2012,Franssen2015} we demonstrate evidence in favour of a Wright-Fisher drift model.  We discuss the potential for rapid approximate methods to approximate evolutionary systems under genetic drift.


\section{Results} 

\begin{figure}[H]
\begin{center}
\includegraphics[width=1\textwidth]{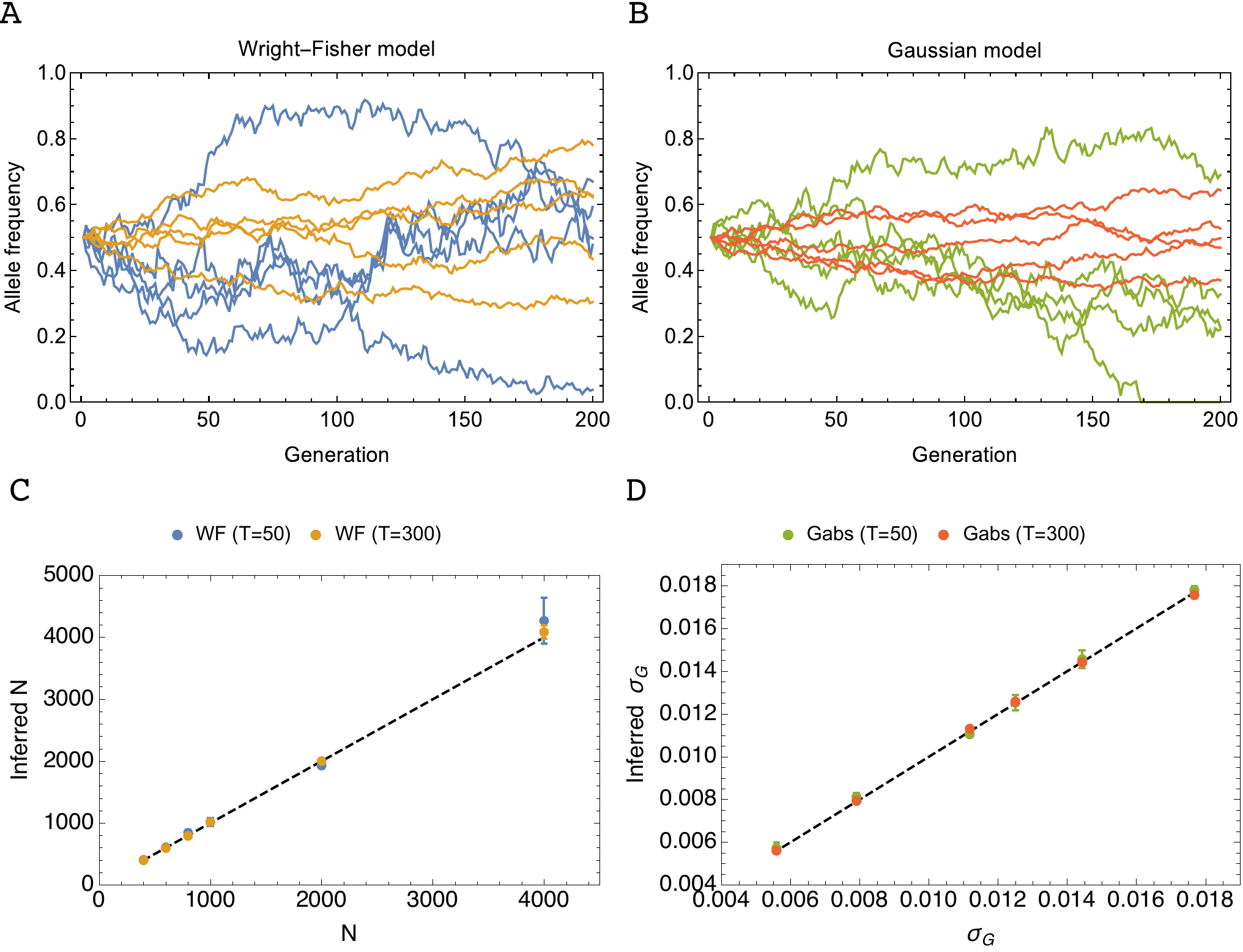}
\caption{\label{fig:Fig_EstNsim_WFandG} {\bf Wright-Fisher and Gaussian models of allele frequency propagation} (A) Example trajectories generated under a Wright-Fisher model with population sizes $N=400$ (blue) and $N=4000$ (yellow). (B) Example trajectories generated under a model of Gaussian diffusion with $\sigma_{G}=0.018$ (green) and $\sigma_{G}=0.006$ (red).  (C) Inferred versus simulated population sizes given observations over $T=50$ and $T=300$ generations of simulated data generated with exact Wright-Fisher propagation. (D) Inferred $\sigma_{G}$ vs simulated $\sigma_{G}$ for equivalent calculations using the Gaussian model.  Simulations used for inference were generated with sampling depth $C=100$, sampling period  $\Delta t=10$, and starting frequency $q(0)=0.5$. Error bars are calculated across three replicate calculations.}
\end{center}
\end{figure}

Given sufficient data, our implemented drift models correctly inferred the underlying population size (or variance) from simulated data describing the respective diffusion process (see Fig.~\ref{fig:Fig_EstNsim_WFandG}).  At large population sizes (or smaller variances) the expected rate of change in an allele frequency due to dispersion declines, so that a longer period of observation was required to estimate $N$ (or $\sigma_{G}$) to a high level of accuracy.  However, accurate estimates were obtained from all simulated populations given 300 generations of sampling (see also Supporting Text on the effect of the number of trajectories on inferred parameters).

\begin{figure}[H]
\includegraphics[width=1\textwidth]{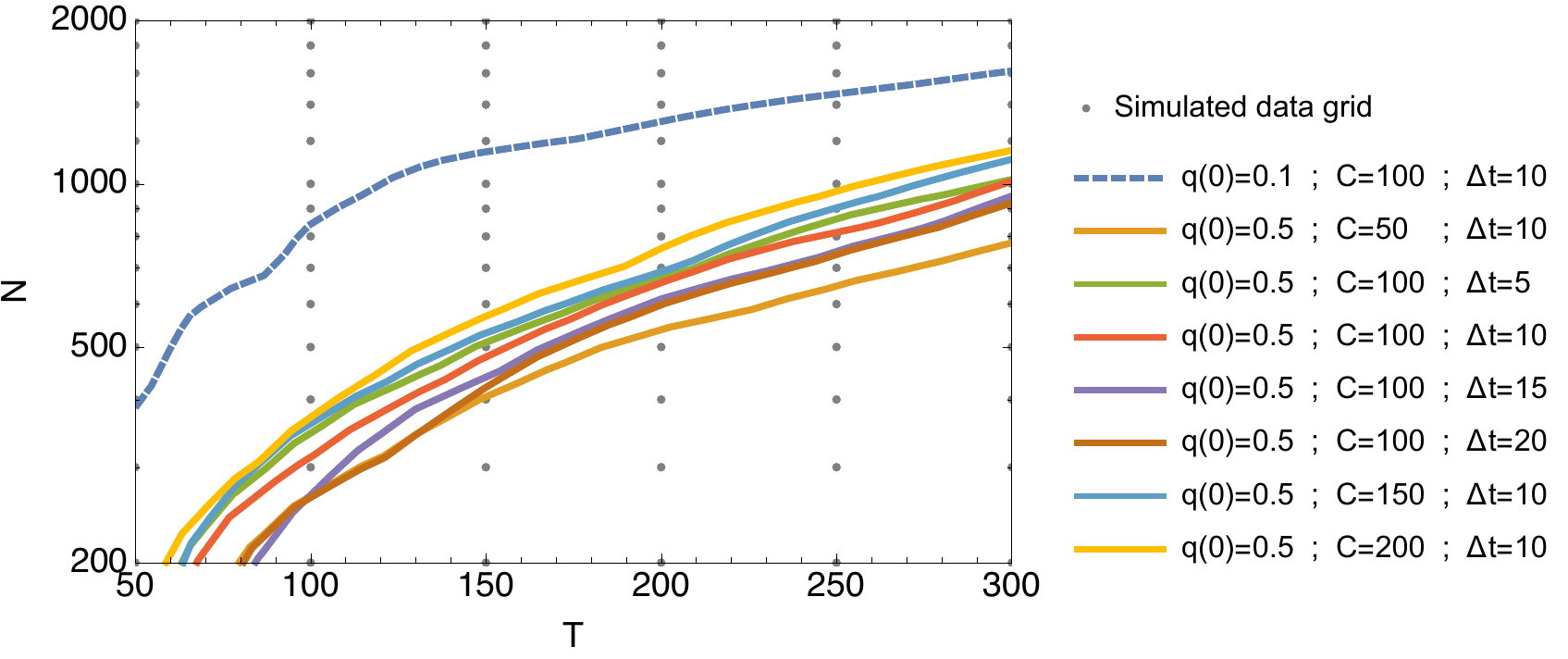}
\caption{\label{fig:Fig_NvsDurationvsDeltaL} {\bf Potential to identify a Wright-Fisher model of evolution.} Contours show lines of constant likelihood difference $\Delta L$ per locus per sampling instant by population size $N$ and experimental duration $T$, between the exact Wright-Fisher and Gaussian drift models when data is generated by Wright-Fisher propagation. Each contour represents the threshold below which correct model identification is possible at comparable likelihood differences.  Solid lines show the contour $\Delta L=0.01$; a dashed line shows the contour $\Delta L=0.05$, for each set of parameters.  Contours were generated by interpolation of data generated at specific combinations of population size and experimental duration, shown as gray dots, and smoothing with an exponential moving average.}
\end{figure}

Given simulated allele frequency data generated according to a Wright-Fisher model without selection, discrimination between the Wright-Fisher and Gaussian models of drift was not always possible; a threshold time, sometimes of 300 generations or more, was required for correct model identification (Fig.~\ref{fig:Fig_NvsDurationvsDeltaL}).  The underlying population size of the system, $N$, was a critical factor in determining the length of the threshold required; at higher $N$, the change via drift is insufficient for model discrimination.  Further factors influenced the threshold; for example trajectories starting at lower frequencies were more informative of the drift model due to increased frequency dependence and the importance of higher moments characterizing the Wright-Fisher model at the frequency boundaries.  An increased depth and frequency of sampling increased the extent of information available for inference; each improved the ability for model discrimination (see Supporting Text).

The presence of natural selection acting within the population led to errors in the estimation of $N$, but did not compromise identification of the Wright-Fisher model (see Supporting Text).  Selection leads to systematic changes in allele frequency with time, and consequentially an increased allele frequency variance.  As such, introducing selection at subsets of loci in the simulated data caused an underestimation of $N$ (under a neutral model) in proportion to the magnitude of selection, and the proportion of loci at which selection acted.  However, the correct inference of a Wright-Fisher model in each case suggests that selection does not confound correct model identification.

\begin{figure}[H]
\includegraphics[width=1\textwidth]{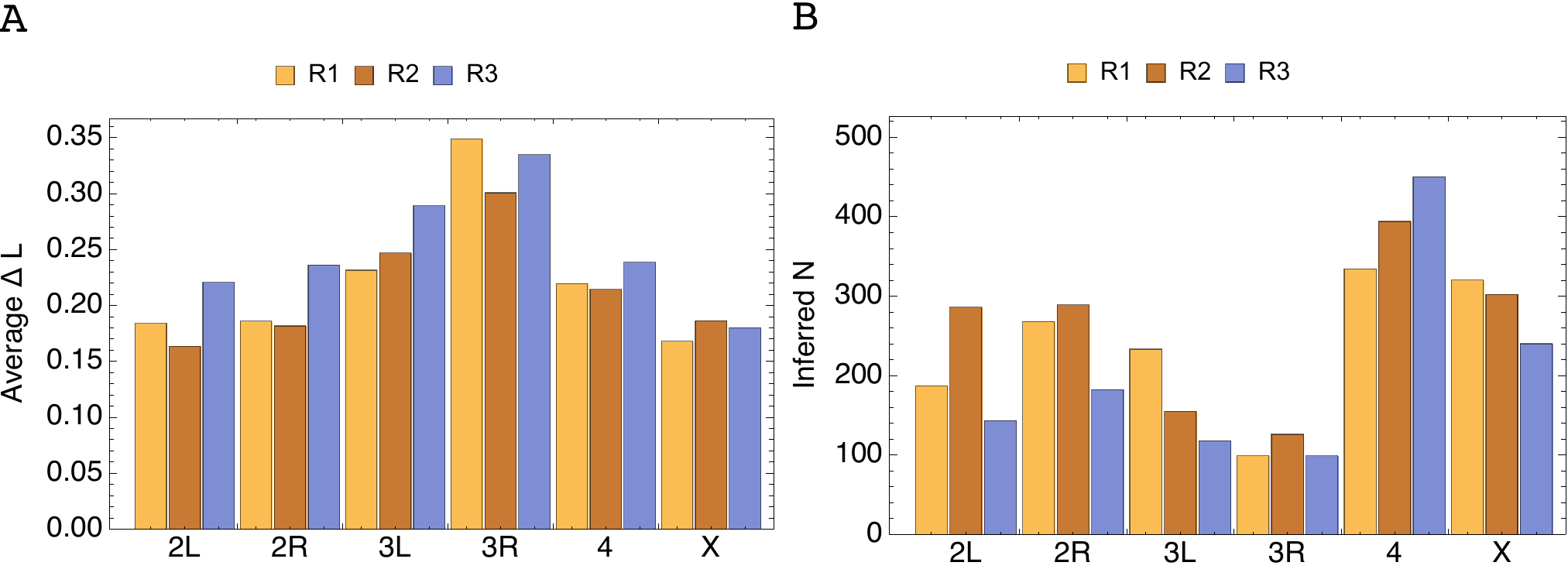}
\caption{\label{fig:Fig_NandLLHDataAllLociNEWDATA} {\bf Population size estimates from Drosophila experimental evolution time-series \cite{Franssen2015} and average likelihood per locus, between exact Wright-Fisher and Gaussian propagation with absorbing boundaries.}}
\end{figure}

Application of both drift models to genomic data from an evolutionary experiment \cite{Franssen2015} showed a better likelihood fit for the Wright-Fisher, as opposed to the Gaussian diffusion model, across the dataset (Fig.~\ref{fig:Fig_NandLLHDataAllLociNEWDATA} (B)).  While an improved fit was not seen for the Wright-Fisher model across all statistical measures considered (see Supporting Text), a clear result in favour of this model was seen via a likelihood calculation.  Estimated population sizes calculated under the Wright-Fisher model are shown in Fig.~\ref{fig:Fig_NandLLHDataAllLociNEWDATA} (A).  Consistent with the identification of selection in the data (\cite{Franssen2015}) these estimates are lower than the reported consensus size of 1000. In addition, a similar calculation was done for the subset of loci in all chromosomes that did not reach fixation. This intended to verify if the success of improved performance associated with Wright-Fisher model did not come from fixation events being naturally included in this drift model, as opposed to artificially modelled, as is the case of the Gaussian drift model used here. The tendency across chromosomes observed in Fig.~\ref{fig:Fig_NandLLHDataAllLociNEWDATA} was not altered, although the average differences in likelihood obtained were slightly lower.

\section{Discussion}

The Wright-Fisher model is the most popular discrete-time approach for modelling populations, describing a population as a succession of randomly drawn, non-overlapping generations in a population of constant size.  However, evaluation of the explicit model is computationally intensive, requiring repeated matrix multiplications for each evaluation.  For this reason, published approaches for inferring selection within a population of finite size have utilised a variety of approximations to the Wright-Fisher model when accounting for genetic drift.\\

Here we have considered the extent to which a Wright-Fisher model is possible to infer from time-resolved allele frequency data.  Applied to a large dataset from an evolutionary experiment, we demonstrate that a Wright-Fisher model is identifiable under a likelihood model.  In so far as Wright-Fisher models can be compared to arbitrarily similar drift models, a Wright-Fisher model can never truly be proven to be correct through the analysis of experimental data.  Nevertheless, under the comparison applied here, we have validated a Wright-Fisher model of genetic drift using data from a biological population.\\

Our calculations further show that the identification of Wright-Fisher drift is not trivial, and may not be replicable in other datasets; in situations where the time over which a population is observed is short, where the underlying population size is large, or where sampling is shallow or sparse, Wright-Fisher drift may be indistinguishable from variance in a Gaussian model.  Under such circumstances the potential for the use of alternative, rapid approximations to the Wright-Fisher approach is clear.  The Gaussian model described here provides one such approach, for which an analytical solution is possible; scope remains for research into alternative procedures.\\


\section{Methods} \label{sec:Methods}


\subsection{Simulated data generation}\label{sec:SimDataGen}

Simulations were performed using both a model of Gaussian diffusion on the interval [0,1] with absorbing boundaries, and using a Wright-Fisher model. 

Parameters for simulations were chosen to reflect those relevant to recent Evolve and Resequence ($E\&R$) experiments \cite{OROZCO-terWENGEL2012,Franssen2015} and representative simulation studies \cite{Kofler2014}, including the population size ($N$), initial frequency distribution ($q(0)$), sequencing coverage depth ($C$), experiment length ($T$), sampling period ($\Delta t$), number of replicates and number of loci ($L$) used to infer population parameters.  Simulations considered evolution at a single locus.  In accordance with the intervals of time considered, mutation was neglected.
A binomial sampling process was used to simulate sequencing of the population.  To study the effects of selection on the inference, simulations were generated in which, for either 1\% or 10\% of loci, selection coefficients were sampled from a uniform distribution in the intervals $[-0.01,0.01]$.


\subsection{A continuous state-space HMM for integer data}\label{sec:filterHD}

Inferences of drift parameters were conducted using a continuous state-space Hidden Markov Model (HMM) for one-dimensional integer data, developed from that used in a previous publication~\cite{Fischer2014}. As with traditional approaches involving HMM, it incorporates a dynamical hidden model, $P(q(t_{k})|q(t_{k-1}),\theta)$ and an emission model, $P(D(t_k)|q(t_k))$, where $D_i(t_k)=\{n_i(t_k),C_i(t_k)\}$ describes the number of observations of a specific allele $n_i(t_k)$, and the total read depth $C_i(t_k)$, at generation $t_k$ and for each locus $i$ in a data set.  Here, we assume a binomial emission model, that is:

\begin{equation}
P(D_{i}(t_k)|q_{i}(t_{k}))={{C_{i}(t_k)}\choose {n_{i}(t_{k})}} q_{i}(t_k)^{n_{i}(t_{k})} (1-q_{i}(t_k))^{(C_{i}(t_k)-n_{i}(t_k))}
\end{equation}

Estimation of parameters $\theta$ was achieved via a forward algorithm, consisting of multiple predict-update steps, by combining sampling with a period $\Delta t=t_{k}-t_{k-1}$ generations and propagation $P(q(t_{k})|q(t_{k-1}),\theta)$:

\begin{eqnarray}
P(q_{i}(t_k)|D_{i}(t_{1:k-1}),\theta)= \int dq_i(t_{k-1}) P(q_i(t_k)|q_i(t_{k-1}),\theta)P(q_i(t_{k-1})|D_{i}(t_{1:k-1}),\theta)\label{eq:PredictForward}
\end{eqnarray}

and 

\begin{eqnarray}
P(q_{i}(t_k)|D_{i}(t_{1:k}),\theta)= \frac{P(D_{i}(t_k)|q_{i}(t_k))P(q_{i}(t_k)|D_{i}(t_{1:k-1}),\theta)}{\int dq_{i}(t_k)P(D_{i}(t_k)|q_{i}(t_k))P(q_{i}(t_k)|D_{i}(t_{1:k-1}),\theta)}\label{eq:UpdateForward}
\end{eqnarray}

leading to the likelihood

\begin{eqnarray}
\mathcal{L}(\theta|D)=\sum\limits_{i=1}^{L}\sum\limits_{k}\log \int dq_{i}(t_k)P(D_{i}(t_k)|q_{i}(t_k))P(q_{i}(t_k)|D_{i}(t_{1:k-1}),\theta)\label{eq:Likelihood}
\end{eqnarray}

Optimisation of this likelihood gave an estimate of the drift parameter $\theta$.

\subsubsection{Transition matrix construction}\label{sec:TransMatConstruct}

Within the above framework, models representing both Gaussian and Wright-Fisher variation were implemented.  The transition probability density matrix for the Gaussian drift model, $P(q(t_{k+1})|q(t_k),\sigma_{G})$, representing frequency evolution between sampling instants $t_k$ and $t_{k+1}$ was constructed by using the analytical solution of the Fokker-Planck equation for a system driven purely by noise, that is:

\begin{equation}
\frac{\partial P(q,t)}{\partial t}=\frac{1}{2}\frac{\partial^{2} P(q,t)}{\partial q^{2}}
\end{equation}

As the normal distribution is a continuous function in the frequency domain, the features associated with the Wright-Fisher at the boundary, namely absorption, are not represented naturally. In order to add this aspect in the Gaussian transition function, we also include absorbing boundaries according to:

\[P_{G_{abs}}(q(t_{k+1})|q(t_k),\sigma_{G})=\left\{
\begin{array}{l}
\mathcal{N}(q(t_{k+1})-q(t_k)| \sigma \sqrt{\Delta t},q(t_k))~: Cond.  \\ 
		\Pi_{0}(t_{k})~:~q(t_k) \neq 0,1~\wedge~q(t_{k+1}) = 0\\ 
		\Pi_{1}(t_{k})~:~q(t_k) \neq 0,1~\wedge~q(t_{k+1}) = 1\\
		1~:~q(t_{k})={0,1}
\end{array} 
\right.
\]

and

\begin{equation}
\begin{split}
Cond.&=~q(t_{k}) \neq {0,1}~\wedge~q(t_{k})-3\sigma\sqrt{\Delta_{k+1}} < q(t_{k+1}) < q(t_{k})+3\sigma\sqrt{\Delta t}~\wedge~q(t_{k+1})\neq 0,1\\ 
\Pi_{0}(t_{k})&=\int ^{0}_{q(t_{k})-3 \sigma \sqrt{\Delta t}} \mathcal{N}(q(t_{k+1})-q(t_{k})| \sigma \sqrt{\Delta t},q(t_{k}))\\
\Pi_{1}(t_{k})&=\int ^{q(t_{k})+3 \sigma \sqrt{\Delta t}}_{1} \mathcal{N}(q(t_{k+1})-q(t_{k})| \sigma \sqrt{\Delta t},q(t_{k}))\\
\end{split}
\end{equation}

Other approaches based on modelling the behaviour near the absorbing boundaries via beta distributions and spikes \cite{Tataru2015} have also been proven to be a valid approach, and could also be implemented within the HMM model presented above.

Frequency transitions were modelled on an evenly spaced discrete frequency grid on the interval $[0,1]$, with resolution $\frac{1}{400}$.\\

For the exact Wright-Fisher propagation model, $P(q(t_{k+1})|q(t_{k}),N)$, no tractable analytical formulation exists allowing immediate computation at any generation $t_{k}$ \cite{Ewens2012}. The exact transition matrix between $t_k$ and $t_{k+1}$ was therefore found by exponentiation of the one-generation $2N$ by $2N$ transition matrix,

\begin{equation}
P(q(t_{k+1})|q(t_k),N)=P(q(1)|q(0),N)^{\Delta t}
\end{equation}

where $P(q(1)|q(0),N)$ is defined by 

\begin{equation}
P_{i,j}(q(1)|q(0),N)={{2N}\choose {2N \times q_{j}(1)}} q_{i}(0)^{(2N \times q_{j}(1))} (1-q_{i}(0))^{(2N(1-q_{j}(1)))}
\end{equation}

with $i,j=1, ..., 401$.  For values of $N$ smaller or greater than $400$, the inverse distance method was used to interpolate between the nearest points on the discrete binomial distribution. 

In the construction of the propagator matrix we do not make any extra assumptions such as a one-step process on the propagation grid as was the case in \cite{Malaspinas2012}; this simplification forces the Markov chain, represented in the transition matrix, to be restricted to diagonal and off-diagonal matrix entries $P_{i,i+1}$ and $P_{i,i-1}$ \cite{VanKampen1992}. Instead, we calculate the full transition matrix for a specific starting frequency involving all entries. 

\section{Code availability}

The code used for matrix exponentiation and likelihood minimization is available at ($https://github.com/DriftModelSelection$). Pre-computed Wright-Fisher transition matrices for population size above 1000 and frequency grid size of 400 are also available at the same address.

\section{Acknowledgements}\label{Acknowl}

This work was supported by a Sir Henry Dale Fellowship, jointly funded by the Wellcome Trust and the Royal Society [Grant Number 101239/Z/13/Z].



  \bibliographystyle{elsarticle-num} 
  \bibliography{DriftModel_May2016_NRN_WithSuppl.bib}

\begin{thebibliography}{10}
\expandafter\ifx\csname url\endcsname\relax
  \def\url#1{\texttt{#1}}\fi
\expandafter\ifx\csname urlprefix\endcsname\relax\def\urlprefix{URL }\fi
\expandafter\ifx\csname href\endcsname\relax
  \def\href#1#2{#2} \def\path#1{#1}\fi

\bibitem{Barrick2013}
J.~E. Barrick, R.~E. Lenski, Genome dynamics during experimental evolution,
  Nature Reviews Genetics 14~(12) (2013) 827--839.
\newblock \href {http://dx.doi.org/10.1038/nrg3564}
  {\path{doi:10.1038/nrg3564}}.

\bibitem{Culleton2005}
R.~Culleton, A.~Martinelli, P.~Hunt, R.~Carter, Linkage group selection: rapid
  gene discovery in malaria parasites, Genome research 15~(1) (2005) 92--97.
\newblock \href {http://dx.doi.org/10.1101/gr.2866205}
  {\path{doi:10.1101/gr.2866205}}.

\bibitem{Mancera2008}
E.~Mancera, R.~Bourgon, A.~Brozzi, W.~Huber, L.~M. Steinmetz, High-resolution
  mapping of meiotic crossovers and non-crossovers in yeast, Nature 454~(7203)
  (2008) 479--485.
\newblock \href {http://dx.doi.org/10.1038/nature07135}
  {\path{doi:10.1038/nature07135}}.

\bibitem{Bergstrom2014}
A.~Bergstr{\"o}m, J.~T. Simpson, F.~Salinas, B.~BarrŽ, L.~Parts, A.~Zia, A.~N.
  Nguyen~Ba, A.~M. Moses, E.~J. Louis, V.~Mustonen, J.~Warringer, R.~Durbin,
  G.~Liti, A high-definition view of functional genetic variation from natural
  yeast genomes, Molecular Biology and Evolution 31~(4) (2014) 872--888.
\newblock \href {http://dx.doi.org/10.1093/molbev/msu037}
  {\path{doi:10.1093/molbev/msu037}}.

\bibitem{Schloetterer2014}
C.~Schl{\"o}tterer, R.~Tobler, R.~Kofler, V.~Nolte, Sequencing pools of
  individuals [mdash] mining genome-wide polymorphism data without big funding,
  Nature Reviews Genetics 15~(11) (2014) 749--763.
\newblock \href {http://dx.doi.org/10.1038/nrg3803}
  {\path{doi:10.1038/nrg3803}}.

\bibitem{Hill:1966fe}
W.~G. Hill, A.~Robertson, {The effect of linkage on limits to artificial
  selection.}, Genetics Research 8~(3) (1966) 269--294.
\newblock \href {http://dx.doi.org/10.1017/S001667230800949X}
  {\path{doi:10.1017/S001667230800949X}}.

\bibitem{Illingworth2011}
C.~J.~R. Illingworth, V.~Mustonen, Distinguishing driver and passenger
  mutations in an evolutionary history categorized by interference, Genetics
  189~(3) (2011) 989--1000.
\newblock \href {http://dx.doi.org/10.1534/genetics.111.133975}
  {\path{doi:10.1534/genetics.111.133975}}.

\bibitem{Jorde2007}
P.~E. Jorde, N.~Ryman, Unbiased estimator for genetic drift and effective
  population size, Genetics 177~(2) (2007) 927--935.
\newblock \href {http://dx.doi.org/10.1534/genetics.107.075481}
  {\path{doi:10.1534/genetics.107.075481}}.

\bibitem{Charlesworth2009}
B.~Charlesworth, Effective population size and patterns of molecular evolution
  and variation, Nature Reviews Genetics 10~(3) (2009) 195--205.
\newblock \href {http://dx.doi.org/10.1038/nrg2526}
  {\path{doi:10.1038/nrg2526}}.

\bibitem{Jonas2016}
{\'A}.~J{\'o}n{\'a}s, T.~Taus, C.~Kosiol, C.~Schl{\"o}tterer, A.~Futschik,
  Estimating the effective population size from temporal allele frequency
  changes in experimental evolution, Genetics 204~(2) (2016) 723--735.
\newblock \href
  {http://arxiv.org/abs/http://www.genetics.org/content/204/2/723.full.pdf}
  {\path{arXiv:http://www.genetics.org/content/204/2/723.full.pdf}}, \href
  {http://dx.doi.org/10.1534/genetics.116.191197}
  {\path{doi:10.1534/genetics.116.191197}}.

\bibitem{Ewens2012}
W.~J. Ewens, Mathematical Population Genetics 1: Theoretical Introduction,
  Vol.~27, Springer Science \& Business Media, 2012.

\bibitem{Waxman2011}
D.~Waxman, Comparison and content of the wrightÐfisher model of random genetic
  drift, the diffusion approximation, and an intermediate model, Journal of
  Theoretical Biology 269~(1) (2011) 79 -- 87.
\newblock \href {http://dx.doi.org/10.1016/j.jtbi.2010.10.014}
  {\path{doi:10.1016/j.jtbi.2010.10.014}}.

\bibitem{Feder2014}
A.~F. Feder, S.~Kryazhimskiy, J.~B. Plotkin, Identifying signatures of
  selection in genetic time series, Genetics 196~(2) (2014) 509--522.
\newblock \href {http://dx.doi.org/10.1534/genetics.113.158220}
  {\path{doi:10.1534/genetics.113.158220}}.

\bibitem{Lacerda2014}
M.~Lacerda, C.~Seoighe, Population genetics inference for
  longitudinally-sampled mutants under strong selection, Genetics 198~(3)
  (2014) 1237--1250.
\newblock \href {http://dx.doi.org/10.1534/genetics.114.167957}
  {\path{doi:10.1534/genetics.114.167957}}.

\bibitem{Terhorst2015}
J.~Terhorst, C.~Schl{\"o}tterer, Y.~S. Song, Multi-locus analysis of genomic
  time series data from experimental evolution, PLoS Genet 11~(4) (2015) 1--29.
\newblock \href {http://dx.doi.org/10.1371/journal.pgen.1005069}
  {\path{doi:10.1371/journal.pgen.1005069}}.

\bibitem{Topa2015}
H.~Topa, {\'A}.~J{\'o}n{\'a}s, R.~Kofler, C.~Kosiol, A.~Honkela, Gaussian
  process test for high-throughput sequencing time series: application to
  experimental evolution, Bioinformatics 31~(11) (2015) 1762--1770.
\newblock \href {http://dx.doi.org/10.1093/bioinformatics/btv014}
  {\path{doi:10.1093/bioinformatics/btv014}}.

\bibitem{Bollback2008}
J.~P. Bollback, T.~L. York, R.~Nielsen, Estimation of 2nes from temporal allele
  frequency data, Genetics 179~(1) (2008) 497--502.
\newblock \href {http://dx.doi.org/10.1534/genetics.107.085019}
  {\path{doi:10.1534/genetics.107.085019}}.

\bibitem{Song2012}
Y.~S. Song, M.~Steinr{\"u}cken, A simple method for finding explicit analytic
  transition densities of diffusion processes with general diploid selection,
  Genetics 190~(3) (2012) 1117--1129.
\newblock \href {http://dx.doi.org/10.1534/genetics.111.136929}
  {\path{doi:10.1534/genetics.111.136929}}.

\bibitem{Steinruecken2014}
M.~Steinr{\"u}cken, A.~Bhaskar, Y.~S. Song, A novel spectral method for
  inferring general diploid selection from time series genetic data, The annals
  of applied statistics 8~(4) (2014) 2203.

\bibitem{Foll2015}
M.~Foll, H.~Shim, J.~D. Jensen, Wfabc: a wrightÐfisher abc-based approach for
  inferring effective population sizes and selection coefficients from
  time-sampled data, Molecular Ecology Resources 15~(1) (2015) 87--98.
\newblock \href {http://dx.doi.org/10.1111/1755-0998.12280}
  {\path{doi:10.1111/1755-0998.12280}}.

\bibitem{Malaspinas2016}
A.-S. Malaspinas, Methods to characterize selective sweeps using time serial
  samples: an ancient dna perspective, Molecular ecology 25~(1) (2016) 24--41.
\newblock \href {http://dx.doi.org/10.1111/mec.13492}
  {\path{doi:10.1111/mec.13492}}.

\bibitem{Der2011}
R.~Der, C.~L. Epstein, J.~B. Plotkin, Generalized population models and the
  nature of genetic drift, Theoretical population biology 80~(2) (2011) 80--99.
\newblock \href {http://dx.doi.org/10.1016/j.tpb.2011.06.004}
  {\path{doi:10.1016/j.tpb.2011.06.004}}.

\bibitem{Buri1956}
P.~Buri, Gene frequency in small populations of mutant drosophila, Evolution
  10~(4) (1956) 367--402.
\newblock \href {http://dx.doi.org/10.2307/2406998}
  {\path{doi:10.2307/2406998}}.

\bibitem{OROZCO-terWENGEL2012}
P.~Orozco-terWengel, M.~Kapun, V.~Nolte, R.~Kofler, T.~Flatt,
  C.~Schl{\o"}tterer, Adaptation of drosophila to a novel laboratory
  environment reveals temporally heterogeneous trajectories of selected
  alleles, Molecular Ecology 21~(20) (2012) 4931--4941.
\newblock \href {http://dx.doi.org/10.1111/j.1365-294X.2012.05673.x}
  {\path{doi:10.1111/j.1365-294X.2012.05673.x}}.

\bibitem{Franssen2015}
S.~U. Franssen, V.~Nolte, R.~Tobler, C.~Schl{\"o}tterer, Patterns of linkage
  disequilibrium and long range hitchhiking in evolving experimental drosophila
  melanogaster populations, Molecular biology and evolution 32~(2) (2015)
  495--509.
\newblock \href {http://dx.doi.org/10.1093/molbev/msu320}
  {\path{doi:10.1093/molbev/msu320}}.

\bibitem{Kofler2014}
R.~Kofler, C.~Schl{\"o}tterer, A guide for the design of evolve and
  resequencing studies, Molecular Biology and Evolution 31~(2) (2014) 474--483.
\newblock \href {http://dx.doi.org/10.1093/molbev/mst221}
  {\path{doi:10.1093/molbev/mst221}}.

\bibitem{Fischer2014}
A.~Fischer, I.~V{\'a}zquez-Garc{\'\i}a, C.~J. Illingworth, V.~Mustonen,
  High-definition reconstruction of clonal composition in cancer, Cell reports
  7~(5) (2014) 1740--1752.
\newblock \href {http://dx.doi.org/10.1016/j.celrep.2014.04.055}
  {\path{doi:10.1016/j.celrep.2014.04.055}}.

\bibitem{Tataru2015}
P.~Tataru, T.~Bataillon, A.~Hobolth, Inference under a wright-fisher model
  using an accurate beta approximation, Genetics 201~(3) (2015) 1133--1141.
\newblock \href {http://dx.doi.org/10.1534/genetics.115.179606}
  {\path{doi:10.1534/genetics.115.179606}}.

\bibitem{Malaspinas2012}
A.-S. Malaspinas, O.~Malaspinas, S.~N. Evans, M.~Slatkin, Estimating allele age
  and selection coefficient from time-serial data, Genetics 192~(2) (2012)
  599--607.
\newblock \href {http://dx.doi.org/10.1534/genetics.112.140939}
  {\path{doi:10.1534/genetics.112.140939}}.

\bibitem{VanKampen1992}
N.~G. Van~Kampen, Stochastic processes in physics and chemistry, Vol.~1,
  Elsevier, 1992.

\end{thebibliography}


\begin{thebibliography}{00}
\bibitem{Kofler2014A}
R. Kofler, C. Schl{\"o}tterer, A guide for the design of evolve and resequencing studies, Molecular Biology and Evolution 31 (2) (2014) 474-483. doi:125 10.1093/molbev/mst221.

\bibitem{Terhorst2015A}
J. Terhorst, C. Schl{\"o}tterer, Y. S. Song, Multi-locus analysis of genomic time series data from experimental evolution, PLoS Genet 11 (4) (2015) 1-29. doi:10.1371/journal.pgen.1005069.

\bibitem{Topa2015A}
H. Topa, A. J{\'o}n{\'a}s, R. Kofler, C. Kosiol, A. Honkela, Gaussian process test for high-throughput sequencing time series: application to experimental evolution, Bioinformatics 31 (11) (2015) 1762-1770. doi:10.1093/bioinformatics/btv014.

\bibitem{Zhao2016A}
L. Zhao, M. Lascoux, D. Waxman, An informational transition in conditioned Markov chains: Applied to genetics and evolution, Journal of Theoretical Biology 402 (2016) 158-170. doi:10.1016/j.jtbi.2016.04.021.

\bibitem{Sorensen2004A}
H. {S{\o}rensen}, Parametric inference for diffusion processes observed at discrete points in time: A survey, International Statistical Review/Revue Internationale De Statistique 72 (3) (2004) 337-354.

\bibitem{Fischer2014A}
A. Fischer, I. V{\'a}zquez-Garc{\'\i}a, C. J. Illingworth, V. Mustonen, High-definition reconstruction of clonal composition in cancer, Cell reports 7 (5) (2014) 1740-1752. doi:10.1016/j.celrep.2014.04.055.

\bibitem{Franssen2015A}
S. U. Franssen, V. Nolte, R. Tobler, C. Schl{\"o}tterer, Patterns of linkage disequilibrium and long range hitchhiking in evolving experimental drosophila melanogaster populations, Molecular biology and evolution 32 (2) (2015) 495-509. doi:10.1093/molbev/msu320.
\end{thebibliography}


\renewcommand\thefigure{A\arabic{figure}}  
\setcounter{figure}{0}
\renewcommand\theequation{A\arabic{equation}} 
\setcounter{equation}{0}

\section*{Supporting Text:}

\subsection*{An increased quantity of data improves the inference of population sizes}

\begin{figure}[H]
\includegraphics[width=1\textwidth]{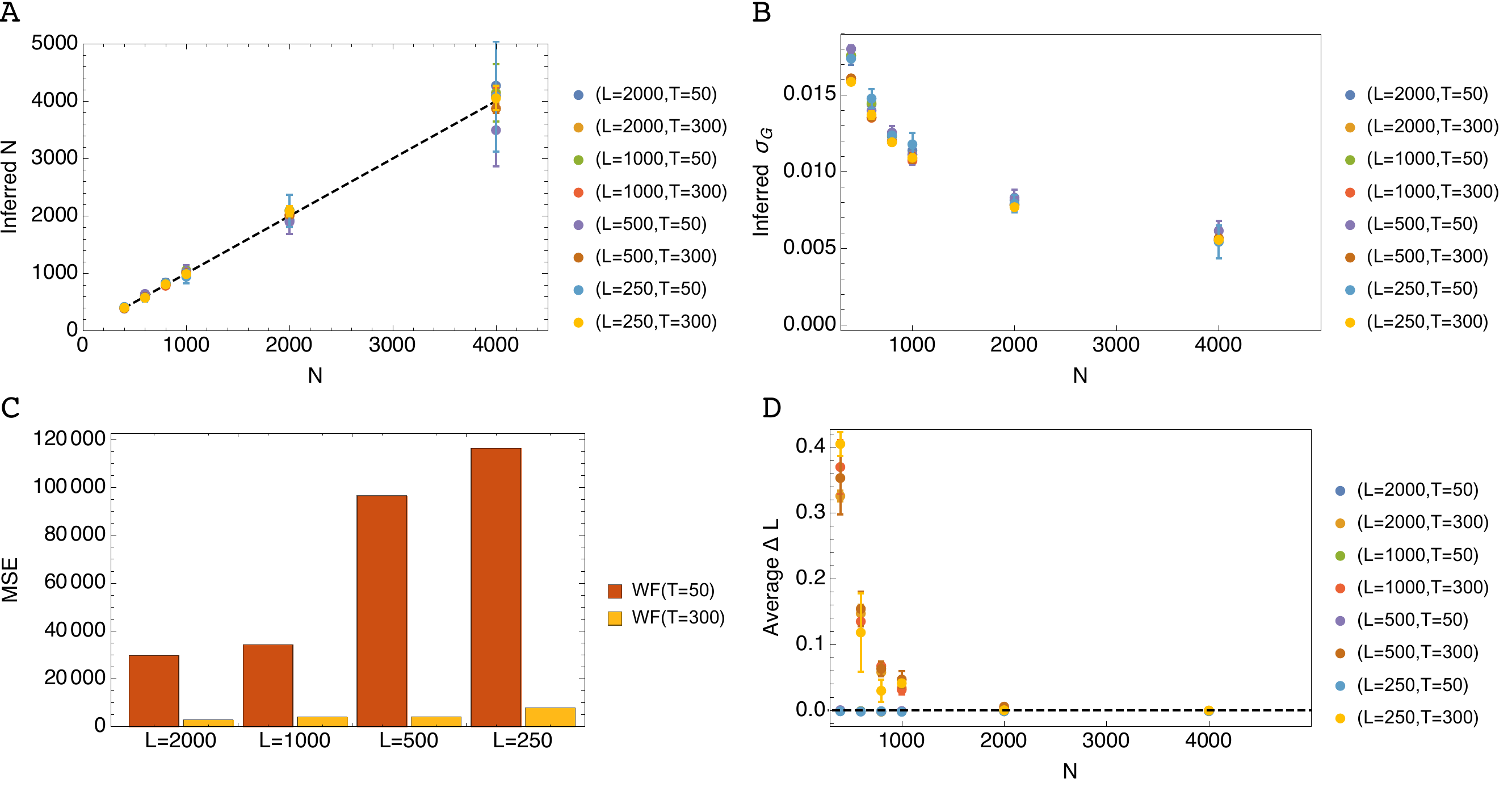}
\caption{\label{fig:Fig_EstNsim_Lsize} {\bf Estimates of population size $N$ and respective performance by length  of evolutionary trajectories ($T$), and size of genomes $L$, for the Wright-Fisher ($WF$) drift model, when simulated data is generated by traditional Wright-Fisher propagation.} (A) Inferred N and (B) Inferred $\sigma_{G}$ vs simulated $N$ for T=50 and 300 generations. (C) Mean-square error between simulated and inferred $N$. (D) Average performance  $\Delta L=L_{WF}-L_{G_{abs}}$ per locus . For all figures sequencing coverage depth $C=100$, sampling period $\Delta t=10$, grid size $400$ and starting frequency $q(0)=0.5$.}
\end{figure}

For the Wright-Fisher model the dispersion of estimates across replicates is larger for larger population sizes due to poor conditioning at these magnitudes, which arises from the variance characteristic of the Wright-Fisher process being of order $O(\frac{1}{N})$ (see Eq.~\ref{eq:Time-dependentWFVar}). This effect and, consequently, the total error in the inferred values, decreases with the number of loci used in the estimates (see Fig.~\ref{fig:Fig_EstNsim_Lsize}). 

\subsection*{Sampling factors affecting the correct inference of Wright-Fisher model parameters}

\begin{figure}[H]
\includegraphics[width=1\textwidth]{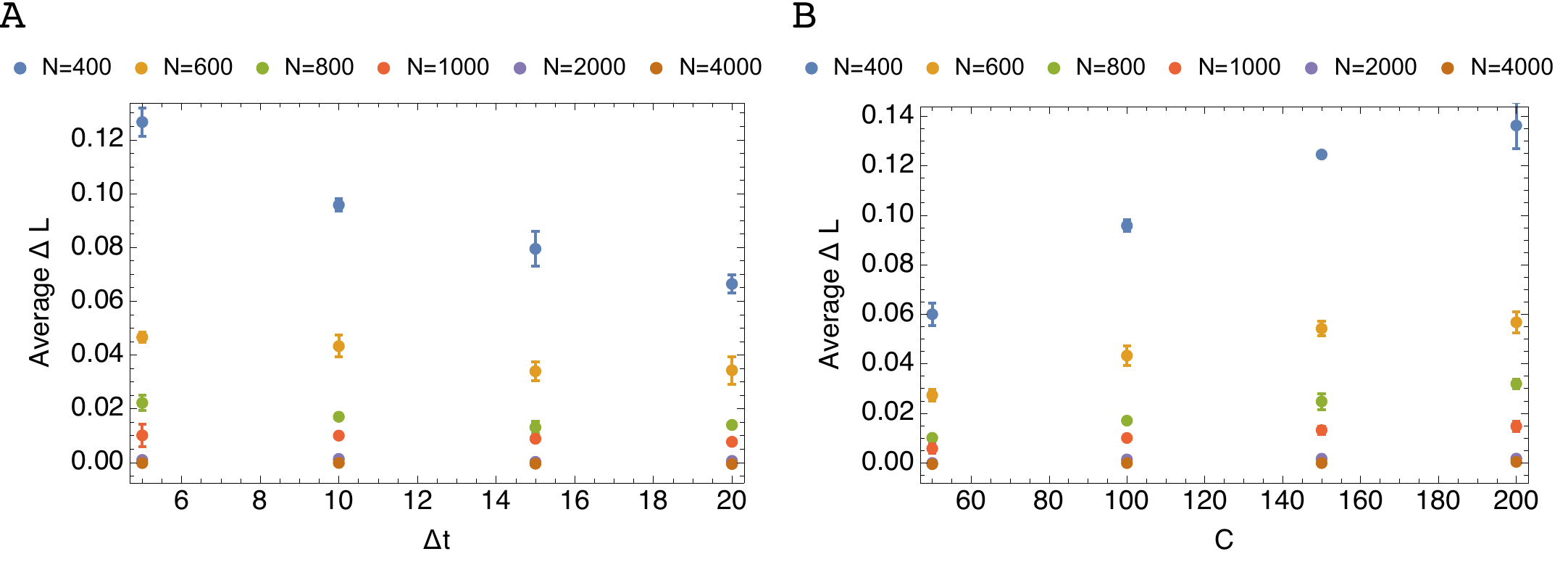}
\caption{\label{fig:Fig_PerfWithSmplFrqAndDpth} {\bf Average performance per locus per sampling instant with sampling period ($\Delta t$) and sequence coverage depth ($C$)}. (A) Average $\Delta L$ by $\Delta t$ with $C=100$. (B) Average $\Delta L$ by $C$ with sampling period $\Delta t=10$.}
\end{figure}

Calculations shown in Fig. 2 of the main text were repeated for different values of $N$, sampling frequency $\Delta t$ and sampling depth $C$.  In each case model inference was performed for simulated Wright-Fisher trajectories at 2000 loci, of length 300 generations, and starting frequency $q(0)=0.5$.  Greater discrimination between models (observed via an increased likelihood for the Wright-Fisher model) was possible given denser sampling of trajectories, and increased sampling depth, as was also clear from observing the threshold curves' order represented in Fig.2 of the main text.  Mean likelihood differences per trajectory and sampling instant are reported in Figure~\ref{fig:Fig_PerfWithSmplFrqAndDpth}.

It was expected that trajectory length ($T$), population size ($N$) and sequencing depth ($C$) would contribute considerably to model identifiability as these parameters have been previously tested in the context of inference of selection \cite{Kofler2014A,Terhorst2015A}. Sampling frequency ($\Delta t$), on the other hand, has not been as extensively explored in the literature of evolutionary time-series analysis, although the importance of having several time-points in conjunction with replicated trajectories is agreed to be fundamental in order to distinguish between selection and drift in relatively small populations \cite{Topa2015A}. Recently, it was reported that for Markov chains such as that represented by the Wright-Fisher process, two observations may not determine entirely the behaviour of the stochastic paths at all intermediate instances \cite{Zhao2016A}, unless the time between these observations is below a characteristic time. This finding is in close proximity to the importance of sampling frequency determined here and, in addition, to the distribution of sampling instances across the duration of the experiment. Outside evolutionary time-series analysis, the importance of how sparse the collection of information is performed has also been proven to be fundamental in correctly inferring parameters of an underlying diffusion process \cite{Sorensen2004A}.  

\subsection*{Effect of natural selection}

\begin{figure}[H]
\includegraphics[width=1\textwidth]{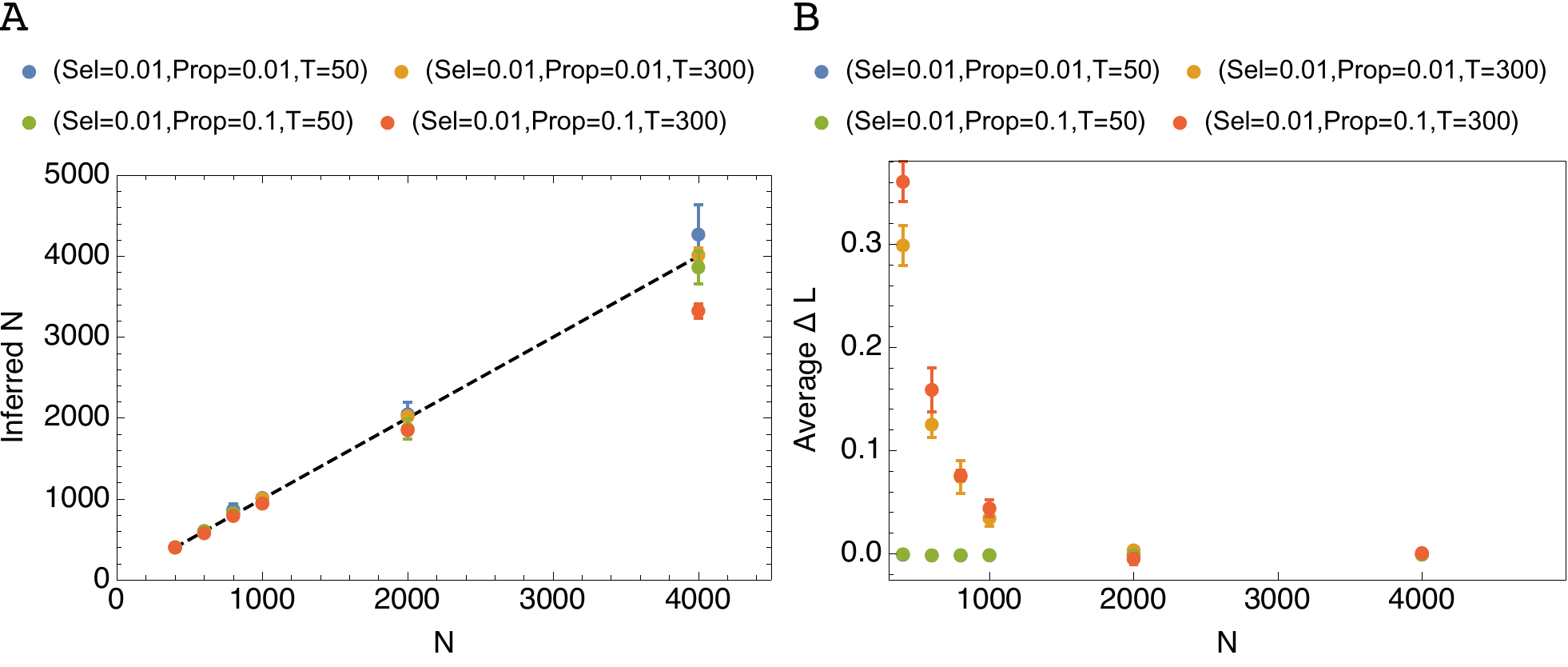}
\caption{\label{fig:Fig_EstNwithSel} {\bf Estimates of drift parameter and respective performance with length  of evolutionary trajectories ($T$), when simulated data is generated by traditional Wright-Fisher propagation with selection.} (A) Inferred $N$  vs simulated $N$, for 300 generations, for several selection strengths and proportion of loci under selection for $q(0)=0.5$, $C=100$ and $\Delta t=100$.  (B) Average performance $\Delta L=L_{WF}-L_{G_{abs}}$ per locus corresponding to (A).}
\end{figure}

The presence of selection increases the variance of the observed allele frequencies, introducing a systematic deviation from the mean.  As such, introducing selection into the simulated data led to an underestimation of $N$, increasing with an increase in the magnitude of selection, and the proportion of loci at which selection acted.  However, correct identification of the Wright-Fisher model was not compromised, a likelihood advantage in favour of this model being inferred in every case (Figure~\ref{fig:Fig_EstNwithSel}).

\subsection*{Alternative measures for evaluating data}

\subsubsection*{Combined forward-backward/predict-update posterior and goodness-of-fit calculation}\label{sec:CombinedHMM}

In addition to the computation of the likelihood we also resorted to another statistic, the goodness-of-fit ($GOF$), taking into account the posterior for each loci frequency at each time-point resulting from the combined forward-backward/predict-update optimization algorithm \cite{Fischer2014A}. \\

As was outlined in the main text, the likelihood function arising from optimization algorithm \cite{Fischer2014A} is

\begin{eqnarray}
\mathcal{L}(\theta|D)=\sum\limits_{i=1}^{L}\sum\limits_{k}\log \int dq_{i}(t_k)P(D_{i}(t_k)|q_{i}(t_k))P(q_{i}(t_k)|D_{i}(t_{1:k-1}),\theta)\label{eq:LikelihoodSuppl}
\end{eqnarray}.

Effectively, $P(q_{i}(t_{k})|D_{i}(1:t_{k-1}),\theta)$, can be determined in an initial step, referred here as the predict step, where we take the data into account.\\

The backward computation is analogous to the forward step described above and the combined forward-backward/predict-update posterior distribution for each loci can be computed by averaging according to Eq.~\ref{eq:CombinedHMM} allowing for all the data to be taken into account, from the initial sampling instant up to the last at $t_{k}=T$ .

\begin{eqnarray}
P(q_{i}(t_{k})|D_{i}(t_{1}:T),\theta)= \frac{P(q_{i}(t_{k})|D_{i}(t_{1}:t_{k}))P(q_{i}(t_{k})|D_{i}(t_{k+1}:T)}{\int dq_{i}(t_{k})P(q_{i}(t_{k})|D_{i}(t_{1}:t_{k}))P(q_{i}(t_{k})|D_{i}(t_{k+1}:T)}\label{eq:CombinedHMM}
\end{eqnarray}

The posterior corresponding to the maximum likelihood estimate can ultimately be used to calculate an additional statistic commonly referred to as Goodness-of-Fit ($GOF$), see Eq.~\ref{eq:GOF}.

\begin{eqnarray}
\mathcal{GOF}(\theta|D)=\sum\limits_{i=1}^{L}\sum\limits_{k}log \int dq_{i}(t_{k})(q_{i}(t_{k})-q^{D}_{i}(t_{k}))^{2}P(q_{i}(t_{k})|D_{i}(t_{1}:T),\theta)\label{eq:GOF}
\end{eqnarray}

Eq.~\ref{eq:GOF} allows us to compute the error, across all loci and sampling instants, in the position of the posterior distribution with respect to the actual data. \\

\begin{figure}[H]
\includegraphics[width=1\textwidth]{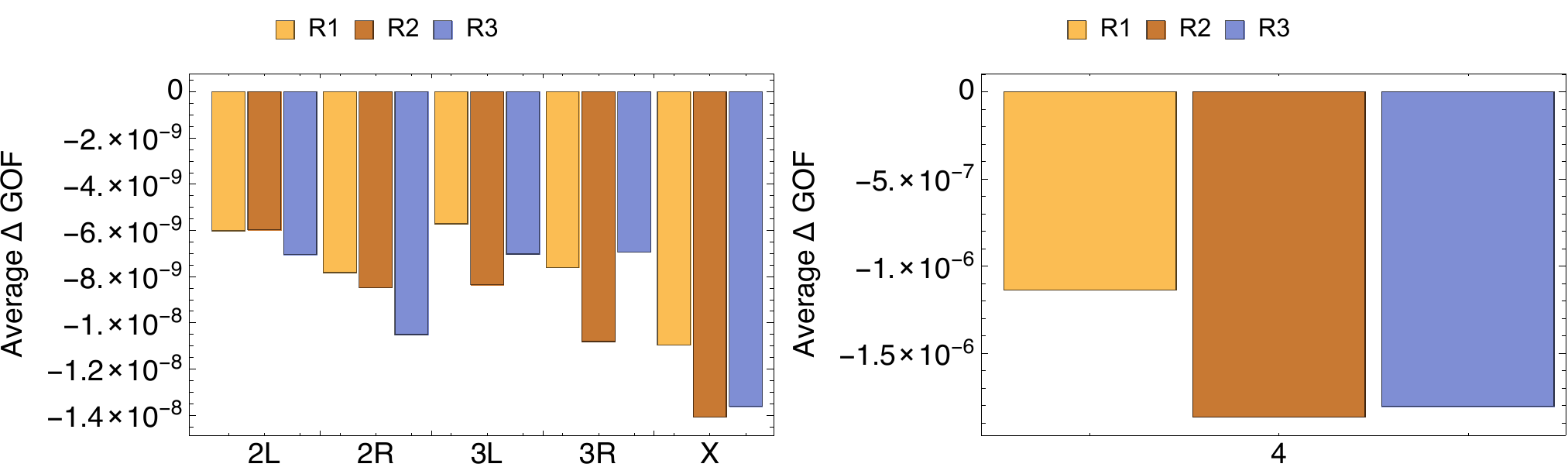}
\caption{\label{fig:Fig_NandGOFDataAllLociNEWDATA} {\bf Goodness-of-fit difference per locus between exact Wright-Fisher and Gaussian propagation models applied to each chromosome of each replicate of the experimental data.}}
\end{figure}

In agreement with the likelihood calculation of the main text, GOF statistics calculated for the experimental data showed a closer fit to the data for the Wright-Fisher, as opposed to the Gaussian model (Figure~\ref{fig:Fig_NandGOFDataAllLociNEWDATA}).

\subsubsection*{Estimation of variance across the frequency spectrum}

We note that, given finite sampling, the Gaussian noise model, in common with the Wright-Fisher model, exhibits frequency-dependent compound variance.\\

Ignoring the effect of the absorbing boundaries, the inherent variance of the Gaussian drift model is frequency-independent and increases linearly with time, as can be derived by applying the law total expectation and total variance:

\begin{equation}
Var^{H}_{G}(t_{k})=\sigma^{2} t_{k}
\end{equation}
while the inherent expectation is constant
\begin{equation}
E^{H}_{G}(t_{k})=E^{H}_{G}(t_{k-1})=q(0)
\end{equation}

A similar calculation for the Wright-Fisher drift model shows the expected frequency-dependent variance at each sampling time as:

\begin{eqnarray}
& E^{H}_{WF}[q(t_{k})]=q(0)\label{eq:Time-dependentWFMean}\\
& Var^{H}_{WF}[q(t_{k})]=q(0)(1-q(0))\left[ 1-(1-\frac{1}{2N})^{t_{k}} \right]\label{eq:Time-dependentWFVar}
\end{eqnarray}

Applying once again the law of total expectation and variance for the sampling step we can obtain variances of the compound sampling problem, at a generation $t_{k}$, under the HMM chain associated with the likelihood function previously presented in Eq.~\ref{eq:LikelihoodSuppl}:

\begin{equation}
\begin{split}
Var^{S}_{G}(t_{k})&=\frac{E^{S}_{G}[q(t_{k})](1-E^{S}_{G}(q(t_{k}))}{C}+ Var^{H}_{G}[q(t_{k})]\\
&=\frac{q(0)(1-q(0))}{C}+(1-\frac{1}{C})\sigma_{G}^{2} t_{k} \label{eq:SmplGVarCompound}
\end{split}
\end{equation}

\begin{equation}
\begin{split}
Var^{S}_{WF}(t_{k})&=\frac{E^{S}_{WF}[q(t_{k})](1-E^{S}_{WF}(q(t_{k}))}{C}+(1-\frac{1}{C})Var^{H}_{WF}[q(t_{k})]\\
&=q(0)(1-q(0))\left\lbrace \frac{1}{C}+(1-\frac{1}{C})\left[ 1-(1-\frac{1}{2N})^{t_{k}} \right]\right\rbrace \label{eq:SmplWFVarCompound} 
\end{split}
\end{equation}

where $C$ is the sampling depth.

Given this calculation, a study was conducted of the extent to which the frequency-dependent variance observed in the data was reproduced by each model.\\

Considering the experimental data, observed allele frequencies were binned according to the predicted posterior means found for each locus and time-point. Plotting the variance of the allele frequency $q(t_{k+1})$ against the measure $q(t_k)(1-q(t_k))$ allowed us to verify the frequency dependence predicted by each drift model either through the analytical derivations represented in Eqs.~\ref{eq:SmplGVarCompound} and \ref{eq:SmplWFVarCompound}, or through the inferred posterior variances resulting from the combined forward-backward/predict-update HMM algorithm outlined above.

Given these measures, the mean squared error between the observed and inferred variances was calculated across the binned frequencies.  Despite no clear pattern being observed in these statistics for each replicate and chromosome, the Gaussian predicted variance calculated through the posterior outperforms the respective Wright-Fisher posterior model if the difference in mean squared error is summed across time-points and replicates. With respect to the variance calculated by applying the analytical solutions represented in Eqs.~\ref{eq:SmplGVarCompound} and \ref{eq:SmplWFVarCompound}, the opposite result is observed. Overall, the use of the posterior variances improves the inferred values of variance when the Gaussian drift model is used, which points to the advantage, in this case, of taking data into account during the HMM algorithm presented above. The same observation is not clearly verified for the Wright-Fisher model. This result contrasts clearly with that reported in the main text where across all chromosomes and replicates the Wright-Fisher is clearly the most representative.

\begin{figure}[H]
\includegraphics[width=1\textwidth]{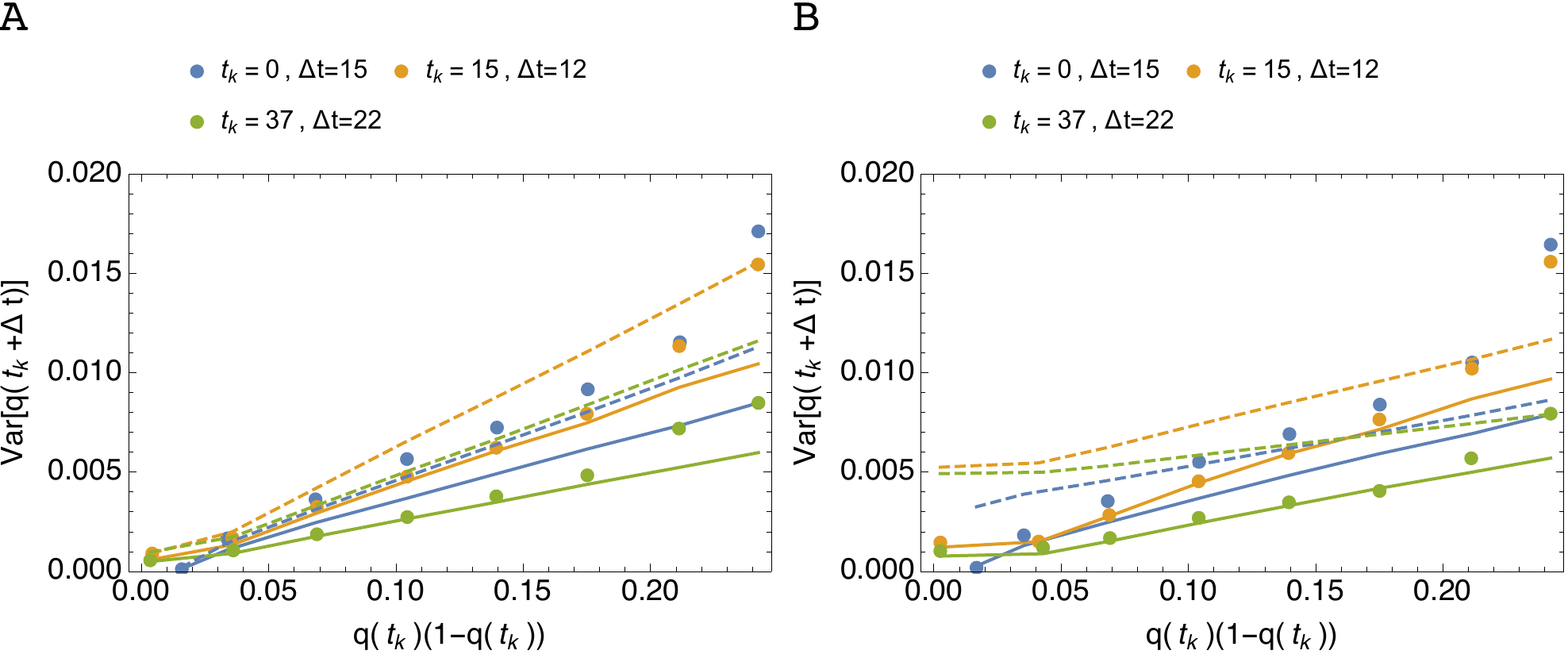}
\caption{\label{fig:Fig_EstVarCompound} {\bf Estimates of compound distribution variance from Drosophila experimental evolution time-series \cite{Franssen2015A} (chromosome $X$, replicate 1).} (A) Compound variance curves obtained with posterior means and variances (Full lines, $WFpost$) as well as with compound variance analytical expressions (Dashed, $WF$) (see Eqs.~\ref{eq:SmplWFVarCompound} and \ref{eq:SmplGVarCompound}) for Wright-Fisher and (B) Gaussian drift models.}
\end{figure}

\begin{figure}[H]
\includegraphics[width=1\textwidth]{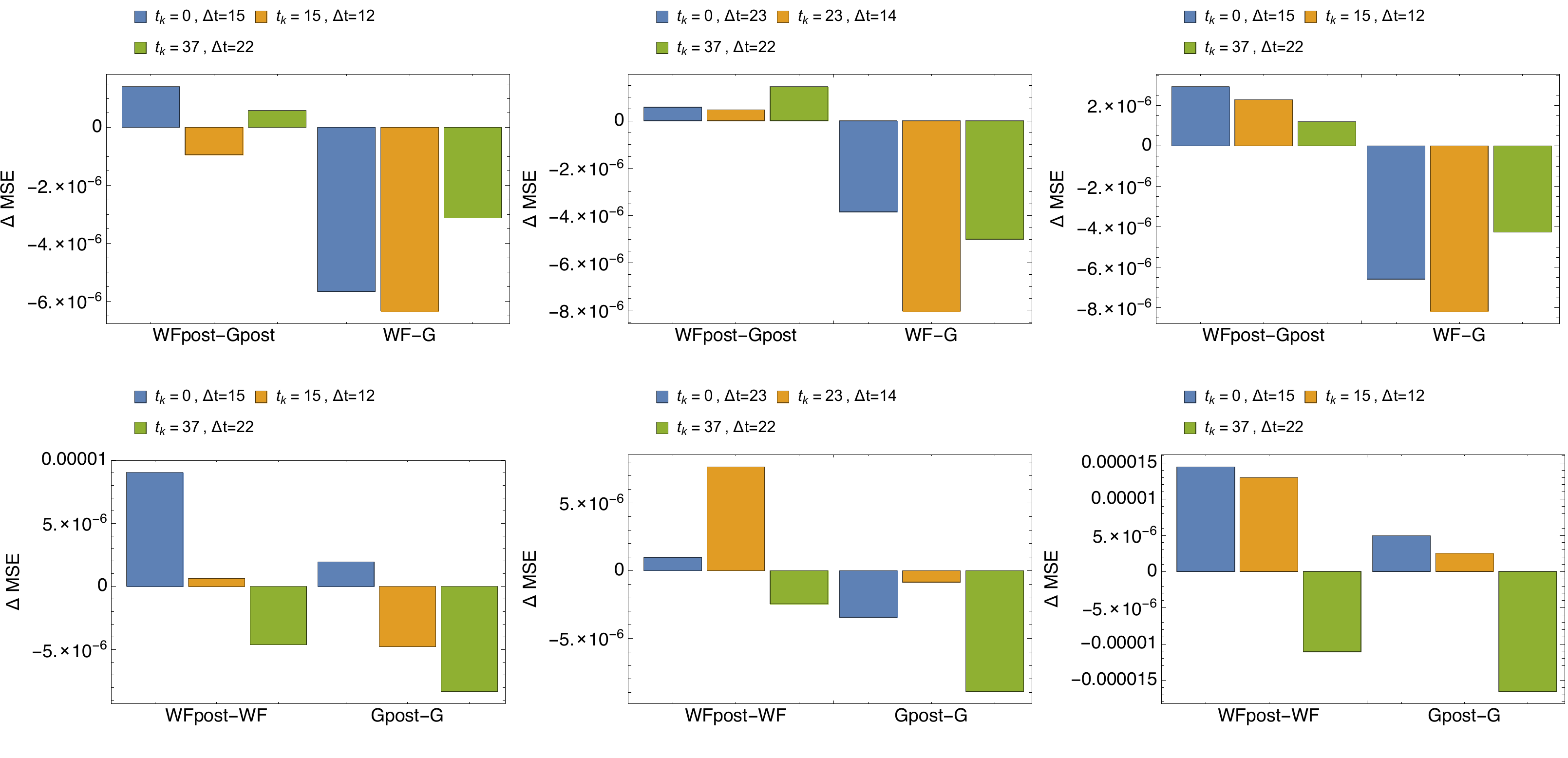}
\caption{\label{fig:Fig_EstVarCompoundBinDistrErrX} {\bf Difference in mean square error in the estimates of compound distribution variance from Drosophila experimental evolution time-series \cite{Franssen2015A} for Wright-Fisher and Gaussian models (chromosome $X$).} $WF_{post},G_{post}$: calculations with posterior variances. $WF,G$: Calculation with analytical solutions. From left to right: replicate $1$, $2$ and $3$.}
\end{figure}

\begin{figure}[H]
\includegraphics[width=1\textwidth]{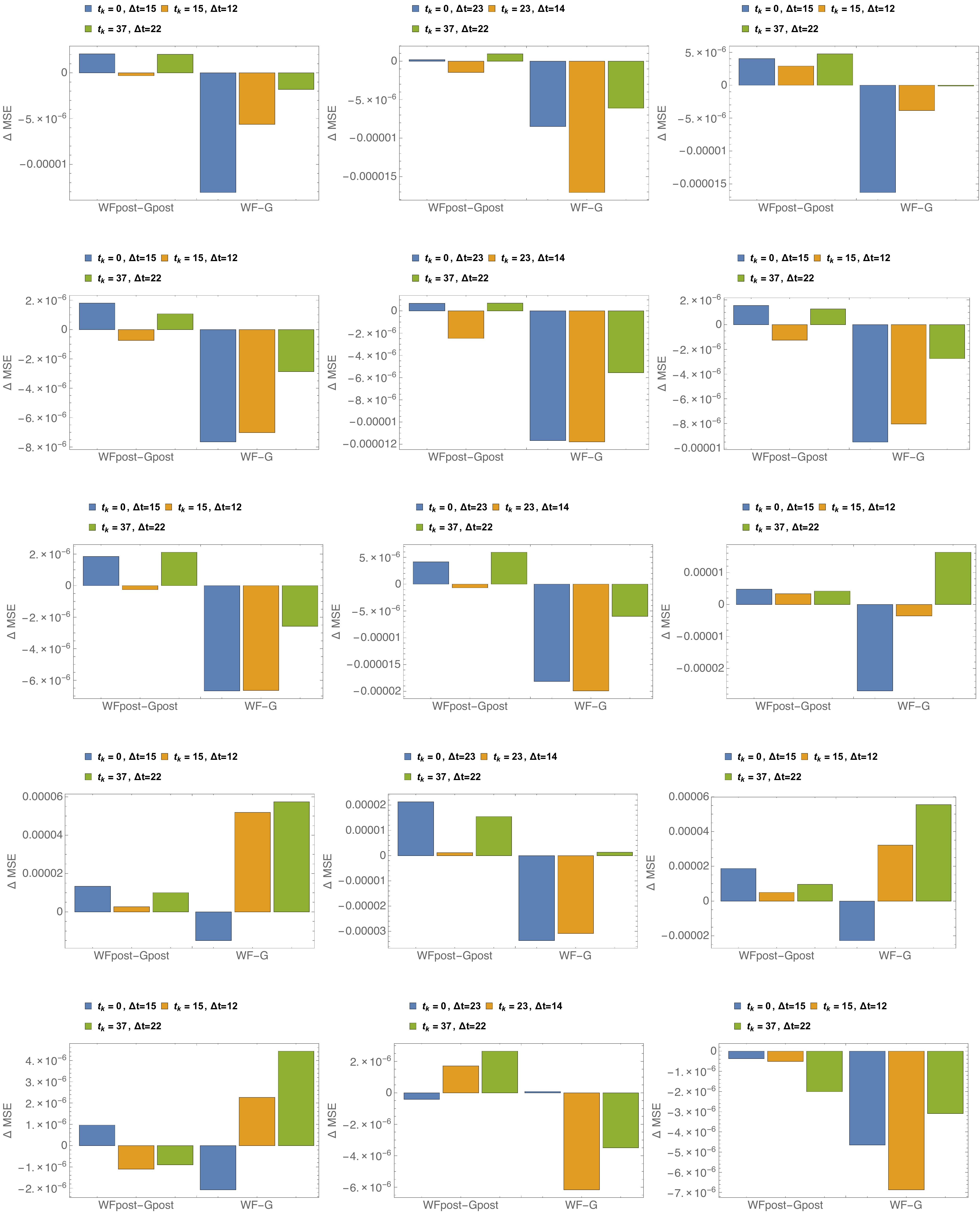}
\caption{\label{fig:Fig_EstVarCompoundBinDistrCrossErr} {\bf Difference in mean square error in the estimates of compound distribution variance from Drosophila experimental evolution time-series \cite{Franssen2015A} for Wright-Fisher and Gaussian models.} $WF_{post},G_{post}$: calculations with posterior variances. $WF,G$: Calculation with analytical solutions. From left to right: replicate $1$, $2$ and $3$. From top to bottom: chromosome $2L, 2R, 3L, 3R, 4$.}
\end{figure}

\begin{figure}[H]
\includegraphics[width=1\textwidth]{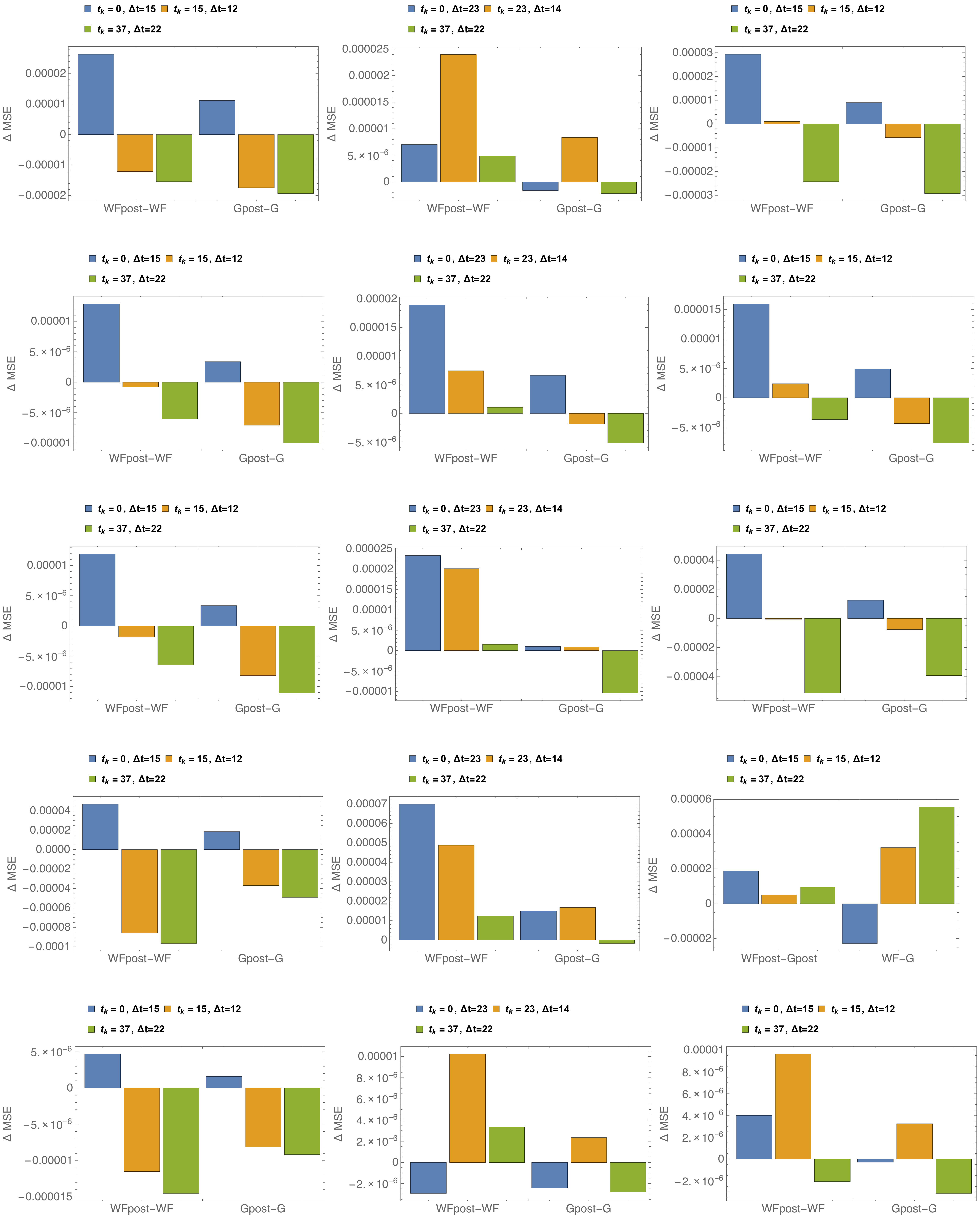}
\caption{\label{fig:Fig_EstVarCompoundBinDistrSameErr} {\bf Difference in mean square error between estimates of compound distribution variance from Drosophila experimental evolution time-series \cite{Franssen2015A} obtained with posterior variances and analytical solutions.} $WF_{post},G_{post}$: calculations with posterior variances. $WF,G$: Calculation with analytical solutions. From left to right: replicate $1$, $2$ and $3$.  From top to bottom: chromosome $2L, 2R, 3L, 3R, 4$.}
\end{figure}

Curves predicted for the X chromosome are shown in Figure~\ref{fig:Fig_EstVarCompound} and the respective error is presented in Fig.~\ref{fig:Fig_EstVarCompoundBinDistrErrX}. Data for other chromosomes is shown in Fig.~\ref{fig:Fig_EstVarCompoundBinDistrCrossErr} and \ref{fig:Fig_EstVarCompoundBinDistrSameErr}

\end{document}